\documentclass[journal]{IEEEtran} 

\usepackage{cite}
\usepackage{hyperref} 
   \usepackage[pdftex]{graphicx} 
\usepackage{amsmath} 
\usepackage{amsfonts}
\usepackage{url} 
\usepackage{color}
\usepackage[normalem]{ulem}
\renewcommand{\d}{\,\textrm{d}}
\renewcommand{\L}{{\cal L}}
\newcommand{\xeq}{{x_\mathrm{eq}}}
\newcommand{\R}{\mathbb{R}}
\newcommand{\E}{\mathbf{E}}

\begin{document}

\title{Simulating the stochastic dynamics and cascade failure of power networks}
%
%
%

\author{
\thanks{Also, Preprint ANL/MCS-P9011-1017, Argonne National Laboratory} Charles Matthews,\thanks{C. Matthews, J. Weare, and M. Anitescu are with the University of Chicago.} \and 
Bradly Stadie,\thanks{B. Stadie is with the University of California, Berkeley.}
Jonathan Weare, 
Mihai Anitescu,\thanks{M. Anitescu is also with Argonne National Laboratory.}
Christopher Demarco \thanks{C. Demarco is with University of Wisconsin, Madison.}
}

%


\maketitle

\begin{abstract}
For large-scale power networks, the failure of  particular transmission lines can offload power to other lines and cause self-protection trips to activate, instigating a cascade of line failures. In extreme cases, this can bring down the entire network. Learning where the vulnerabilities are and the expected timescales for which failures are likely is an active area of research. In this article we present a novel stochastic dynamics model for   a large-scale power network along with a framework for efficient computer simulation of the model including long timescale events such as cascade failure. We build on an existing Hamiltonian formulation and introduce stochastic forcing and damping components to simulate small perturbations to the network.  Our model and simulation framework allow assessment of the particular weaknesses in a power network that make it susceptible to cascade failure, along with the timescales and mechanism for expected failures. 
\end{abstract}

\begin{IEEEkeywords}
Cascading failure, stochastic differential equations, Hamiltonian systems, rare event simulation	
\end{IEEEkeywords}

%

\section{Introduction} \label{sec::intro}

Many  modern infrastructure systems depend on the delivery of electricity through a reliable and robust always-on power network.  However, the stability of a particular network topology under small perturbations or malicious attack can be difficult to accurately predict, particularly when these networks possess self-protection.  Self-protective action is ubiquitous in electric networks and acts to disconnect components when the state trajectory exits an acceptable operating region defined for each component.  Protective action is a positive effect when a small number of components are saved from damage by disconnection and the overall system ``rides through'' in a stable fashion that keeps the entire network energized.  In many scenarios, however, the protective action exacerbates the effect of an initial disturbance and induces a cascading phenomenon that disconnects many components, de-energizing significant portions of the network.  The large number of variables and the combinatoric breadth of cascade scenarios make anticipating the frequency and location of potential problems a challenging task.

The phenomenon described above is typically termed ``cascading failure.'' Such widespread interruption of service may indeed result from misoperation of protective relays \cite{protection_misoperation}, and in these circumstances the terminology of cascading ``failure'' is appropriate.  However,  protective relay actions also may each be locally appropriate, operating in accordance with the relays desired settings, and yet a significant portion of the network is nonetheless de-energized by their action.  This situation reflects an obvious and  inherent tension in selecting thresholds for component disconnection when designing and setting protection.  Equipment owners' objectives are naturally oriented toward protecting components from damage (favoring ``early'' disconnection); in contrast, transmission system operators (TSOs) typically seek to maintain uninterrupted service to the greatest number of customers (favoring ``late'' disconnection).  This tension  has emerged  in recent discussions between generation owners and TSOs  regarding  rate-of-change-of-frequency  relay protection for generators \cite{ROCOF}.  One application of the tools to be developed here will be to better inform this type of trade-off, providing a more global view of the impact of  protective relay settings and network topology on resulting patterns of network outage.  This global perspective on the interaction of protection is vital; one part of the system can be weakened or disconnected, which after load rebalancing could result in another part of the system disconnecting if critical transmission lines or transformers become overloaded and trip. The effect can snowball to de-energize large parts of the network. This type of analysis  can expose hidden fragility in a network: a system may appear robust to small perturbations  (i.e., it possesses a strong margin of stability in small signal stability analysis that does not account for relay action), and yet a small shock in a particular region can trigger a cascade within the network. 

Because of the high impact and visibility of the problem  (the eastern US blackout in New York in 2003 may be viewed as a cascading failure \cite{blackout_report_2003}), cascading network failure has been an extremely active area of research; a survey of the large volume of work through 2011 is provided in \cite{survey_cascading}, and much work has followed.  However, many  articles focus on quasi-steady-state representations of the network, employing Markov chain models to characterize transitions between network configurations, with  transition probabilities  governed by protective relays removing components from service.  Transmission lines have been most widely studied as the class of components removed by relay action.   Enhancements to Markov-based transition approaches seek to capture dynamic effects, with recent work  \cite{Thorp_2017} doing so by constructing a Lyapunov functional ``built on top of" the underlying Markov representation.

The work presented here is built on a nonlinear model representing the electromechanical dynamics of the network.  Probabilistic time variation of load demand is  explicitly represented  to capture continuously acting stochastic disturbances that excite these dynamics.  Protective relay actions are then represented as state-dependent discrete events, disconnecting a component if the state trajectory exits a region defining the associated relay's operational boundary.  The result is a higher-resolution model of network dynamics than those previously considered.
This higher fidelity comes at an additional computational cost, however, and a careful treatment of the numerical solution is required in order to make simulation practical. We therefore introduce both an efficient time-discretization of the model along with a technique that employs multiple parallel simulations of the system to efficiently generate cascade failure events that would otherwise require simulation on extremely long timescales.

The paper is structured as follows. In Section \ref{sec::model} we review the Hamiltonian formulation for the power system model representing electromechanical dynamics, and we introduce the stochastic perturbations to these dynamics;  Section \ref{sec:robustness} introduces the failure mechanism we  employ in our dynamics; Section \ref{sec::simulation} discusses the effective discretization of the dynamics; Section \ref{sec::parrep} describes an advanced sampling technique  that we have found useful in simulating the network. In Section \ref{sec::results} we combine the techniques to simulate a cascade failure in small network; and in Section \ref{sec::conclusion} we briefly summarize our work.

\section{Modeling power network dynamics} \label{sec::model}

We dynamically model an interconnected power network as a graph with $N$ nodes (buses) and $L$ power lines represented as edges between the nodes. Nodes are split into three categories: generator nodes, which supply power to the network; load nodes consuming power; and slack nodes (one per connected component of the graph), whose behavior is fixed by some external source. 

The instantaneous state of the system is given by voltage angle, voltage magnitude, and frequency variables associated with each node, which evolve in time according to a stochastic differential equation (SDE) that we describe below. This equation does not model the very fast timescale dynamics of the switching transients  (i.e.,  no attempt is made to accurately model arc extinction, since an interrupted current flow transitions from hundreds of amperes to zero in a fraction of a second).  Instead, the transition associated with switching action of relays is captured approximately by an abrupt transition in the parameters of the SDE as described in Section \ref{sec:robustness}.

The state of a node $n$ at a time $t$ is described in our model  by its angular velocity $\omega_n(t)$ as well as its voltage magnitude $V_n(t)$ and phase angle $\theta_n(t)$, in normalized ``per unit" coordinates corresponding to the physical power grid. The values of $\theta_n$ and $\omega_n$ are defined relative to a synchronous reference at a time $t$, namely, $\omega_n(t) = \hat{\omega}_n(t) - \hat{\omega}$ and $\theta_n(t) = \hat{\theta}_n(t) - \hat{\omega} t$, where $\hat{\omega}_n(t)$ and $\hat{\theta}_n(t)$ are  the physically observed values at time $t$. The complex voltage phasor at node $n$ is similarly defined relative to a synchronous reference. The phasor is denoted $v_n := V_n \exp(i\theta_n)$, with  imaginary unit $i$ and $|v_n|=V_n$.

The dynamics we impose on the system are inspired by a Hamiltonian dynamics model, whose deterministic form has been previously examined in \cite{Zheng_2010} and \cite{Zheng_2016}.
We first introduce the Hamiltonian
\begin{equation}\label{eqn::Ham}
H(\omega,\theta,V ) := \frac12 \omega^T M \omega  +  \frac12 v^H A^T B_\gamma A v+P \cdot \theta+Q\cdot \ln V
\end{equation}
where $(\ln V)_n = \ln V_n$ and where we denote  the vector of complex voltage phasors as $v$ with its complex conjugate transpose written as $v^H$. The diagonal positive-definite matrix of normalized rotational inertia is denoted $M$, and vectors of real and reactive injected power are denoted $P$ and $Q$, respectively.

We shall treat $P$ and $Q$ as constant vectors in our formulation. In practice, their values will have small fluctuations on short timescales, as well as larger-magnitude periodic  changes from the time of day and the season. Treating these values as constants  simplifies the dynamics and still allows us to consider trajectories on a reasonable timescale (around six hours).

The $N\times N$ nodal admittance matrix $Y_\mathrm{bus} = A^T B_\gamma A $ defines the topology of the network. $A$ is an $L \times N$ incidence matrix defining the connections between nodes, with 
\[
A_{ln} = \left\{\begin{array}{lcl} 1 & & \textrm{If line $l$ enters node $n$ }  \\ -1 & & \textrm{If line $l$ exits node $n$ }\\ 0 & & \textrm{otherwise.} \end{array} \right.
\]
Although we treat the power network as an undirected graph, an implicit sign convention is introduced through the $A$ matrix. $B$ is an $L\times L$ diagonal matrix with entries $(B_\gamma)_{jk}=\delta_{jk} \gamma_j b_k$, where $b_l>0$ gives the susceptance value of line $l$ and $\gamma_l$ is 1 or 0 depending on whether line $l$ is up or down, respectively.  Here $\delta_{jk}$ is the Kronecker delta symbol, $\delta_{jk}=1$ if $j=k$, and $\delta_{jk}=0$ otherwise. 

This Hamiltonian is chosen so that its relevant partial derivatives match the equilibrium power flow equations. Writing the gradient operator as
\begin{equation}
 \nabla = \left[  \nabla_\omega \,,\, \nabla_\theta \,,\, \nabla_V   \right]^T,
\end{equation}
we can compute the partial derivatives of $H$  as
\begin{equation}
 \nabla H = \left[ \begin{array}{c} M \omega \\  - \textrm{Im} ( \textrm{diag}(v^H) A^T B_\gamma A v  ) + P \\ 
		    \textrm{Re}(\exp( \textrm{diag}( -i \theta ) )A^T B_\gamma A v   )  + \textrm{diag}(V)^{-1} Q \end{array} \right].
\end{equation}
Denoting the equilibrium point $x_\mathrm{eq}$ where $\nabla H(x_\mathrm{eq})=\mathbf{0}$ corresponds to the solution of the lossless power flow equations, we have
\begin{align*}
0 &= P_n - \sum_{k=1}^N V_k V_n (Y_\textrm{bus})_{nk} \sin( \theta_n - \theta_k ),\\
0 &= Q_n + \sum_{k=1}^N V_k V_n (Y_\textrm{bus})_{nk} \cos( \theta_n - \theta_k )
\end{align*}
for all $n$.

In Hamiltonian dynamics, the total energy is a first integral, and numerical discretization schemes are constructed with the aim of approximately preserving $H$ \cite{leimkuhler2004simulating}. 
To derive our stochastic dynamics, we begin by assuming that the Hamiltonian dynamics are perturbed by noise in the form of the short-timescale fluctuations in the $P$ and $Q$ vectors. The behavior of the network will then be characterized by the statistics of multiple random trajectories, rather than a single deterministic trajectory.

The dynamics are chosen such that the system remains close to $\xeq$ by perturbing standard Hamiltonian (constant-energy) dynamics by a mild stochastic term representing random fluctuations in the power demands at each node. This amounts to adding two terms into the conservative dynamics: a damping term that slowly drives the system toward $\xeq$ and a noise term  that represents the fluctuations around $P$ and $Q$.
The magnitude of the damping and the noise are small enough  that if we initialize the dynamics close to $\xeq$, then trajectories will likely remain close by, with the fluctuation rate and magnitude of the deviance from $\xeq$ given by parameters in the stochastic process. 

\begin{figure}
\begin{center}
\includegraphics[trim={0.75cm 0 1cm 0},clip,width=0.5\textwidth]{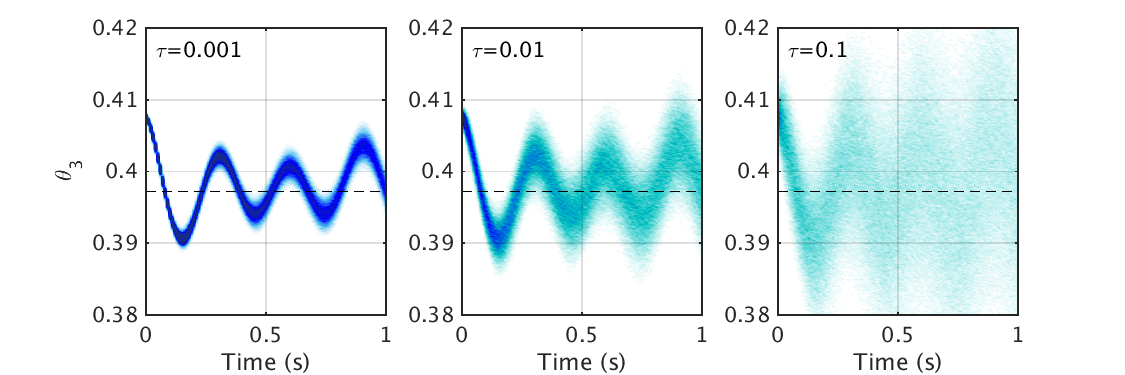}
\caption{We plot the evolving distribution of $\theta_n(t)$ for a single node under the dynamics \eqref{eqn::sdes} with $\epsilon=0.0025$. We initialized the system from a point  away from $\xeq$ at time $0$, with $\xeq$ marked as a black dashed line. As the damping rate is fixed, the variance of the evolving distribution strongly depends on the choice of $\tau$, with the distribution remaining closer to the deterministic trajectory for small $\tau$.  }\label{fig::temp}
\end{center}
\end{figure}

Our proposed dynamics for the network are given by the following  stochastic differential equation, which consists of a conservative Hamiltonian part and a fluctuating Ornstein-Uhlenbeck process \cite{pavliotis2014stochastic}:
\begin{align} \label{eqn::sdes}
 \d x &= J \nabla H(x) \d t - \epsilon S \nabla H(x) \d t + \sqrt{2\epsilon\tau S} \d W_t,
\end{align}
where $x=(\omega,\theta,V)^T$ is the state vector and $W_t \in \R^{3N}$ is a vector of $3N$ independent Wiener processes \cite{pavliotis2014stochastic}. The $\epsilon\geq0$ parameter gives the reciprocal timescale (or damping parameter) for the decorrelation of stochastic fluctuations, and $\tau\geq0$ defines the strength of the noise introduced (akin to a temperature parameter in physical interpretations of the dynamics \cite{leimmatbook} 
). Fixing $\epsilon$ and increasing $\tau$ increase the statistical variance in solutions at a time $t$, increasing the spread away from the deterministic solution (see Figure  \ref{fig::temp}). Too small a value of $\tau$ defeats the purpose of using stochastic methods, while too large a value of $\tau$ gives nonphysical behavior.

$S$ and $J$ are symmetric and skew matrices, respectively, defined as 
\[
J = \left[ \begin{array}{ccc} 0 & -I_{l}-I_g & 0 \\ I_{l}+I_g & 0 & 0 \\ 0 & 0 & 0 \end{array} \right], \quad S = \left[ \begin{array}{ccc} 0 & 0 & 0 \\ 0 & I_{l}+I_g & 0 \\ 0 & 0 &  I_{l} \end{array} \right]
\]
with $(I_{l})_{jk} = \delta_{jk} 1_{l}(j)$ and $(I_{g})_{jk} = \delta_{jk} 1_{g}(j)$, where $1_{l}(j)=1$ if node $j$ is a load node and $1_{g}(j)=1$ if node $j$ is a generator node  (and zero otherwise).    This ensures that the dynamics for the slack nodes, as well as the magnitude of voltage for the generator nodes, are frozen in time.

The $\gamma$ parameters will evolve according to the event-driven dynamics  described in Section \ref{sec:robustness}.  For $\gamma$ fixed (provided some assumptions on the growth and boundedness of $H$), solutions to \eqref{eqn::sdes} are known to be ergodic \cite{ma2015complete} with respect to the invariant measure
\begin{equation} \label{eqn::invdist}
 \mu_\tau(x) = Z_\tau^{-1} \exp(- H(x) / \tau), \quad Z_\tau = \int \exp(- H(x) / \tau) \d x
\end{equation}
provided $\epsilon$ and $\tau$ are both positive. Thus, if $\epsilon\tau>0$, then ergodicity implies  \cite{ma2015complete} that 
\[
\E(f(x)) := \int f(x) \mu_\tau(x)\d x = \lim_{T\to\infty} \frac1T \sum_{t=1}^T f(x(t))
\]
where $x(t)$ is a solution to \eqref{eqn::sdes}. If $\epsilon=0$, then we recover Hamiltonian (constant-energy) dynamics. In contrast, if $\tau=0$ and $\epsilon>0$, then we obtain the steepest decent method when using an Euler discretization \cite{milstein2004stochastic}, with a trajectory converging toward $\xeq$. 

The ergodicity property when $\epsilon\tau>0$ provides a useful way of tuning the $\tau$ parameter to realistic data. As solutions to the SDE are distributed according to $\mu_\tau(x)$, the long-time distribution of one variable can be found by computing the relevant integral to find its expectation. For example, the distribution of $\omega$ can be found by integrating out the remaining variables in the distribution:
\[
\omega \sim \int \int \mu_\tau(\omega,\theta,V) \d \theta \d V \propto \exp\left(-\frac12\omega^TM\omega/\tau\right)
\]
since the Hamiltonian $H(x)$ in $\mu_\tau$ is separable. Thus we expect $\omega \sim \mathrm{Normal}(0, \tau M^{-1} )$.

Physical observations corroborate that the angular momentum is normally distributed \cite{lu2006design}, and we can choose the  parameter $\tau$ to match the physically observed variance of the angular velocity as if $M=\textrm{diag}(m)$ then $\textrm{var}(\omega_n) = \tau/m_n$.
In normal operations, the range for frequency is $59.95$ Hz to $60.05$ for $99.95$ \% of the time \cite{lu2006design}, and thus the standard deviation $\mathrm{std}(\omega)\approx0.016$ Hz. This gives a realistic choice for $\tau/m_n$ as around $2.5\times10^{-4}$.

While changing $\tau$ can have a significant impact on the timescales and qualitative features of the simulation (see Figure \ref{fig::temp}), we find that the results are far more tolerant to different choices of the damping parameter $\epsilon$. One simply needs to choose a timescale long enough (and hence $\epsilon$ small enough) to ensure that the timescale of the fluctuations are longer than important features of the simulation. In what follows, we have chosen $\epsilon=0.05s^{-1}$.

The continuous dynamics themselves are useless without an efficient way to generate solution trajectories on a computer. This is achieved through a novel discretization scheme that reduces observed error in averages, discussed in detail in Section \ref{sec::simulation}. Before addressing this practical concern, however, we introduce in the following section the mechanism  by which we characterize the propagation of line failures in the system.

\section{Line failure and quantifying network robustness} \label{sec:robustness}

Our motivation for introducing the stochastic dynamics, as outlined in Section \ref{sec::intro}, is to gather statistics and classify failures of the network. The most common way to approximate the behavior of the physical network at a failure point  is to mimic the responses of the thermal relay on each power line, by removing a line from service when the energy on the line becomes too large (i.e. setting $\gamma_l\to0$). One can emulate this behavior in simulation in many ways.  Perhaps the simplest is checking the line flows at an equilibrium predicted by the dc power flow approximation $\xeq$ as in \cite{wang2012power}.  
In  \cite{DeMarco_2001}, \cite{Zheng_2010}, and \cite{Zheng_2016}, deterministic equations based on the Hamiltonian \eqref{eqn::Ham} (or similar formulations) are coupled with a continuous dynamics for the $\gamma$ parameters  approximating  the relay decision logic.

Here we choose a simple event-driven dynamics for $\gamma$: we  remove line $l$ by switching $\gamma_l=0$ instantaneously when some condition is met. In our model, lines are either removed intentionally as part of some scenario (replicating some environmental effects) or in order to model thermal and stability constraints imposed by the maximum load allowed on a line. In the latter case we monitor the line energy of line $l$, defined as
\[
\Theta_l := \frac12 b_l \left|v_{l_\text{in}} - v_{l_\text{out}} \right|^2 ,
\]
where ${l_\text{in}}$ and ${l_\text{out}}$ are the indexes of the nodes that line $l$ joins in the network.

We model the failure dynamics as an irreversible change that alters network topology and hence moves the equilibrium point $\xeq$. An interesting feature of this formalism is that we can capture the timescale and trajectory of the system as it transitions between neighborhoods of $\xeq^{\mathrm{(old)}}$ to $\xeq^{\mathrm{(new)}}$, which correspond to the stable regions before and after the line failure.

Our general algorithm for the simulation of the network is as follows.
\begin{enumerate}
 \item Initialize the system by setting $x\leftarrow\xeq$ for the initial choice of $\gamma$.
 \item Evolve the system for time $h$ using the numerical discretization scheme, and set $t\leftarrow t+h$. \label{alg::start}
 \item Any line $l$ with line energy $\Theta_l$ above a given threshold has $\gamma_l\leftarrow0$. 
 \item Any line $l$ no longer connected to a slack node  has {${\gamma_l\leftarrow0}$}.
 \item If $t<T$, then go to \ref{alg::start}; otherwise finish.
\end{enumerate}

\begin{figure}
\begin{center}
\includegraphics[trim={0cm 0 0cm 0},clip,width=0.5\textwidth]{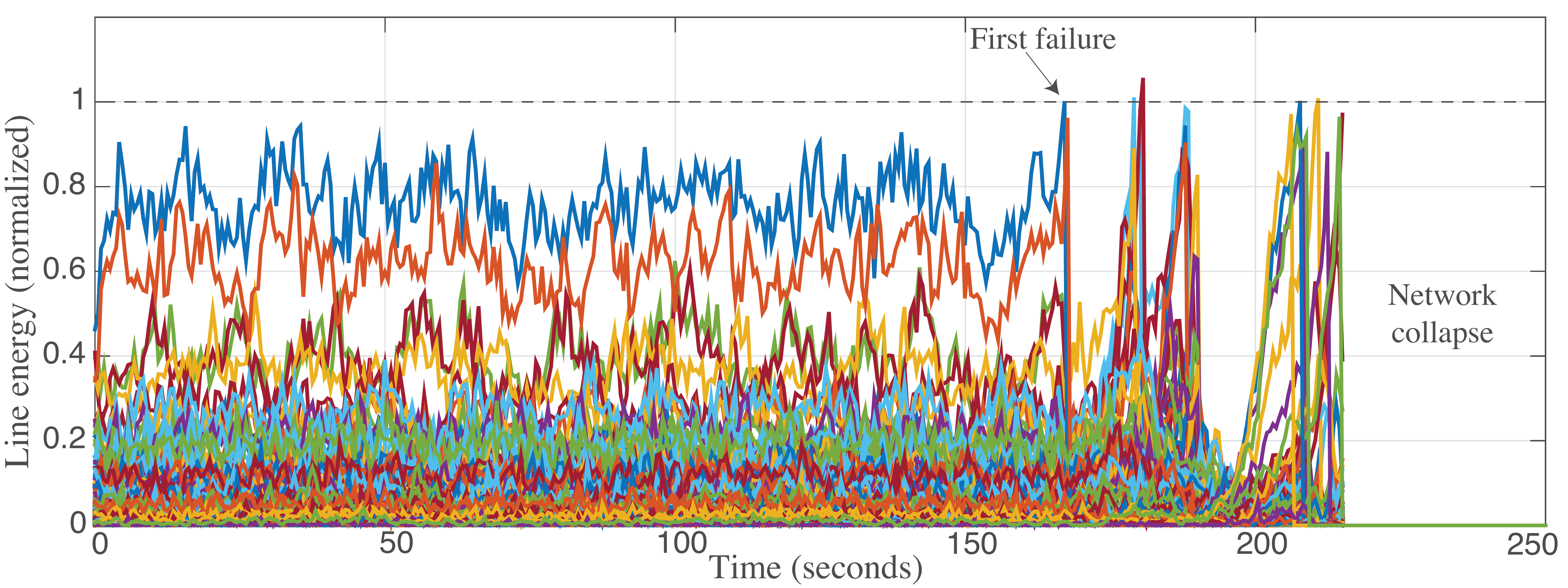}
\caption{We simulate the network plotted in Figure \ref{fig::network} using the algorithm given in Section \ref{sec::simulation}. The model is weakened initially by removing a critical line, and we use a low threshold for the thermal relays, ensuring rapid failures. Plotted is the line energy, normalized by the tripping threshold, for all 453 lines. Random fluctuations cause the first line to disconnect after 164 seconds, rapidly leading to total failure of the network. 
}\label{fig::le}
\end{center}
\end{figure}

We would expect that, just as in the real system, no lines should fail when we run a simulation initialized with all lines up ($\gamma={\mathbf 1}$) and with the system in the neighborhood of the equilibrium point $\xeq$. For a weakened system, however, with one or two lines already down, the stress on some lines may be large enough to cause them to fail either immediately or within some time horizon. Note that this failure may occur when the system transitions between the neighborhoods around the successive equilibria corresponding to a change in $\gamma$; and while the equilibrium points themselves may not have lines with excessive line energies, the probable transition path between them may. 
An example is shown in Figure \ref{fig::le}, where an initial failure causes a cascade of follow-on events as the system transitions between configurations.

With our framework presented in Section \ref{sec::model} we can compute the likelihood of a failure occurring in a particular scenario, and we can give the expected time and transition pathway for the failures themselves. We do so by generating a large number of realizations of the solutions to \eqref{eqn::sdes} and computing average statistics. 


We use the total supplied load as a measure for the performance of the network. The load served by the network is a function of which lines are up or down, with
\[
L(\gamma) := \frac{P^T I_l \Lambda(\gamma)}{P^T I_l   \mathbf{1} }, 
\]
where  $\Lambda(\gamma)\in\R^N$ is the vector indicating which nodes are connected to a reference node, with
\[
 \Lambda_n(\gamma) = \left\{ \begin{array}{lcl} 1 &    & \textrm{Node $n$ is connected to a slack node} \\ 0 & & \textrm{otherwise}. \end{array} \right.
\]
For a trajectory $\gamma(t)$, $L$ is piecewise constant in time, with changes occurring only when $\gamma(t)$ is modified such that a load bus is removed from a subnetwork containing the slack node. Of key interest for a trajectory is the normalized cumulative load served over an interval $[0,T]$
\[
 \L(T) := \frac1T \int_0^T L(\gamma(t)) \d t \in [0,1], \quad \L(0)=L(\gamma(0)).
\]
This quantity gives a reference of the load that has been shed by the network due to lines being removed. Thus,  $\L(t)=1$ indicates that the network has supplied the total demanded load for time $t$. Since we assume that lines cannot be restored during a simulation, $\L$ will be strictly decreasing after the first line failure.  This is a more valuable statistic than the number of lines remaining or buses connected, since lines can fail or islands can form without consequence to the overall utility of the network. 

A simple strategy is to test scenarios of interest and look at the distribution of  $\L(T)$ for some time horizon $T$.  Simulations with a low cumulative load served reveal a weakness in the network that can be studied by using trajectory data.

Of particular interest is establishing a typical order of events to quantify a network's progress toward failure. The most natural way to characterize this progress is to look at the order that lines trip in simulations.  A large network, however,  may have a very large number of relatively likely sequences of line failures, due to the combinatorial explosion in sequences of lines of a certain length.

We have found that a simple and effective way to characterize the path toward failure is to use a clustering algorithm that groups lines together and to classify the failure path by the order that lines in those groups are removed from service. We use Matlab's \texttt{linkage} command to build hierarchical clusters of downed lines 
using the distance metric between lines $j$ and $k$, $d_{jk}$, defined as
\[
d_{jk} := \E( | \rho(\gamma_j(t)) - \rho(\gamma_k(t)) |\,\vert\,\rho(\gamma_j(t))\rho(\gamma_k(t))<\infty ),
\]
where
\[
\rho(\gamma_l(t)) = \min( \{t\,:\, t>0,\,\gamma_l(t)=0 \} )
\]
gives the failure time of line $l$ for a trajectory $\gamma(t)$. Thus we may think of $d_{jk}$ as being the mean difference in failure times between lines $j$ and $k$, conditional on their failing. Lines that often fail together have a shorter distance in this metric and are consequently grouped together by the clustering scheme. The hierarchical scheme we use groups lines into clusters and subclusters to better understand how lines fail together.

\section{Simulation of the network} \label{sec::simulation}

\begin{figure}
\begin{center}
\includegraphics[width=0.3\textwidth]{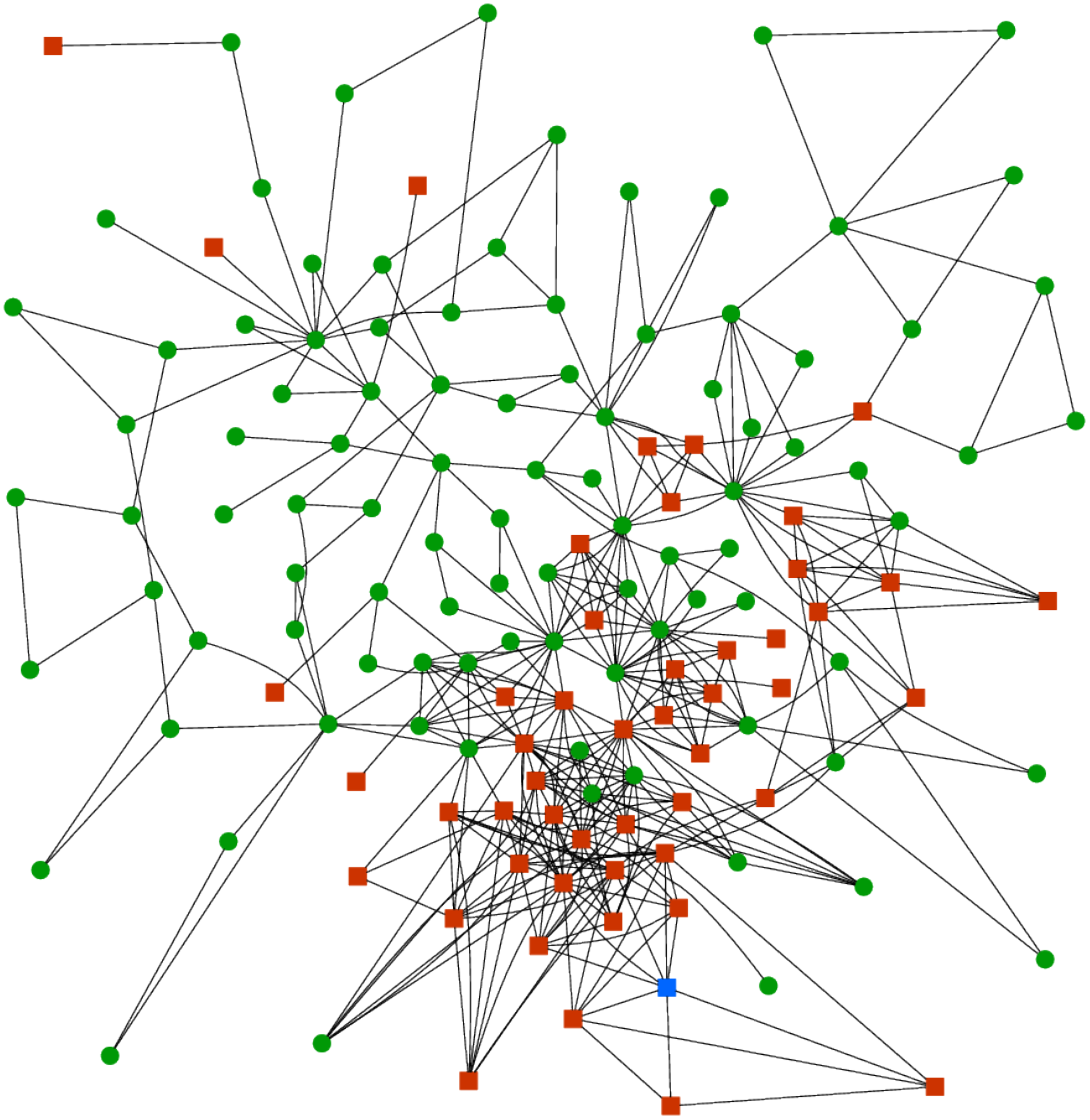}
\caption{The  benchmark power network \cite{Chow1995} used in our tests. There are 145 buses (95 load buses as green circles, 49 generators as red squares, and 1 slack bus as a blue square) connected by 453 power lines. }\label{fig::network}
\end{center}
\end{figure}

Since the goal of our modeling is to probe the statistics of the system as it evolves, we seek to generate a large number of trajectories by solving the dynamics \eqref{eqn::sdes}. Because of the dimensionality of the state $x$ and the complexity of the Hamiltonian defining the dynamics, finding analytical solutions is not feasible. We turn instead to a numerical timestepping algorithm with time discretization parameter (timestep) $h>0$ to advance from $x(t)$ to $x(t+h)$.   The parameter $h$ is usually chosen after some experimentation---too small a choice increases the computational effort required to simulate to time $T$ (the required wall time is usually $O(T/h)$), while increasing $h$ can create instability in the system or a large error in observed averages. 

We seek to minimize two quantities when considering algorithms for propagating the dynamics: computational cost and discretization error.   The algorithms we consider involve simple matrix-vector operations, with the major cost coming from the evaluation of $\nabla H$. Although solutions to the dynamics \eqref{eqn::sdes} will sample the distribution $\mu_\tau$, the discretization algorithms give an approximation to the solution with a statistical error that vanishes as $h\to0$  (as long as the algorithm is consistent). 
This discretization error we are interested in is the observed bias in average quantities taken over the discretized trajectory. 

We compare results for four discretization algorithms implemented in a python code, on a small benchmark system of 145 buses and 453 lines taken from \cite{Chow1995}, plotted in Figure \ref{fig::network}. The computational cost (measured in wall time) and introduced discretization bias (using the computed error in averages)  are compared at a fixed stepsize $h=2^{-5}$s and time window $T=100$s, initializing the variables at $\xeq$ with all lines up ($\gamma=\mathbf{1}$).  More details about the schemes can be found in the Appendix.


\begin{table}[bt]
\renewcommand{\arraystretch}{1.3}
\caption{Comparison between schemes}
\label{tab::comparison}
\centering
\begin{tabular}{c||c|c|c}
\hline
  Scheme &   Wall Time & Line Energy Error  & Hitting Time Error \\
\hline\hline
Euler & $\mathbf{8.46}$s & $34.1\pm0.8\%$ & $66.4\pm0.5\%$\\
Heun & $13.94$s & $5.2\pm0.7\%$ & $23.8\pm1.8\%$\\
Structure Pres. & $16.41$s & $1.7\pm0.5\%$ & $4.1\pm2.0\%$\\
L-M & $8.70$s & $\mathbf{0.8\pm0.5}\%$ & $\mathbf{2.8\pm1.4\%}$\\
\hline
\end{tabular}
\end{table}

The results of the experiment are summarized in Table \ref{tab::comparison}. We give the mean wall-clock time for one run of each of the schemes. We also give the relative error in the average variance of the line energy for one line and the average time for the line's energy to reach $4.5\%$ of its susceptance value. These quantities should be general enough to be representative of the quality of other averages.

In our test the LM scheme is significantly more accurate than the Heun method and costs roughly half the time of the structure-preserving scheme.   Hence in what follows we use the LM scheme to compute statistical quantities and to propagate solutions from time $t$ to time $t+h$.

\section{Parallel Replica Dynamics } \label{sec::parrep}
We can trivially parallelize our simulation method to gather statistics efficiently by running multiple independent simulations in parallel. We cannot easily parallelize the dynamics in time, however, so increasing available computing resources will generally not allow running the system over a longer time horizon for a given wall-clock limit. We may utilize parallelism in our computation of the $\nabla H$ term to speed each iteration; however, this is not without difficulty and seldom leads to a linear speedup.  

The Parallel Replica Dynamics (ParRep) technique, originally proposed for rare-event simulation in computational chemistry \cite{Voter1998,Binder2015595}, allows one to achieve an (approximately) linear speedup in the number of processors while still recovering exact statistics. Multiple copies (replicas) of the simulation are run simultaneously until one of them has a line failure. The other simulations are discarded, and all simulations are restarted from the replica which had the failure (after some decorrelation time).  The correct, exponentially distributed, failure times are then calculated by scaling the first exit times by the number of replicas used.

Full details of the algorithm can be found in the given references or in the Appendix.

\section{Numerical tests} \label{sec::results}
We perform two tests on the network plotted in Figure \ref{fig::network}, with  145 buses (including 49 generators and 1 slack bus) and 453 lines. 

\subsection{Low-threshold network} \label{sec::results1}
We set our threshold parameter for all lines to be reasonably low, so that a line is disconnected if its energy reaches $6.5\%$ of its susceptance value. 
For each line $l$, we run ten pilot simulations for $T=600 s$, with the system initialized at $\xeq$ but with $\gamma_l(1)=0$; that is,  line $l$ is disconnected after an initial burn-in time of $1 s$. The average cumulative load served $\L(T)$ was computed for each experiment and compared across all runs. One line in particular ($l=164$) showed a significant deficiency in cumulative served load following its disconnection, even though by disconnecting this one line we do not directly isolate any nodes. In what follows, we perform multiple experiments with this line initially removed.

We run 1,024 simulations for a total time of $T=2$ hours or until the network reaches total failure, using a timestep of $h=0.02$ s with the LM algorithm. This timestep was chosen for a balance between efficiency and accuracy; although the trajectories appeared stable even up to $h=0.04$ s, the statistical properties of the simulation were qualitatively changed when using such a large discretization (we observed that the rate of failure was significantly increased).

\begin{figure}
\begin{center}
\includegraphics[trim={0cm 0 0cm 0},clip,width=0.35\textwidth]{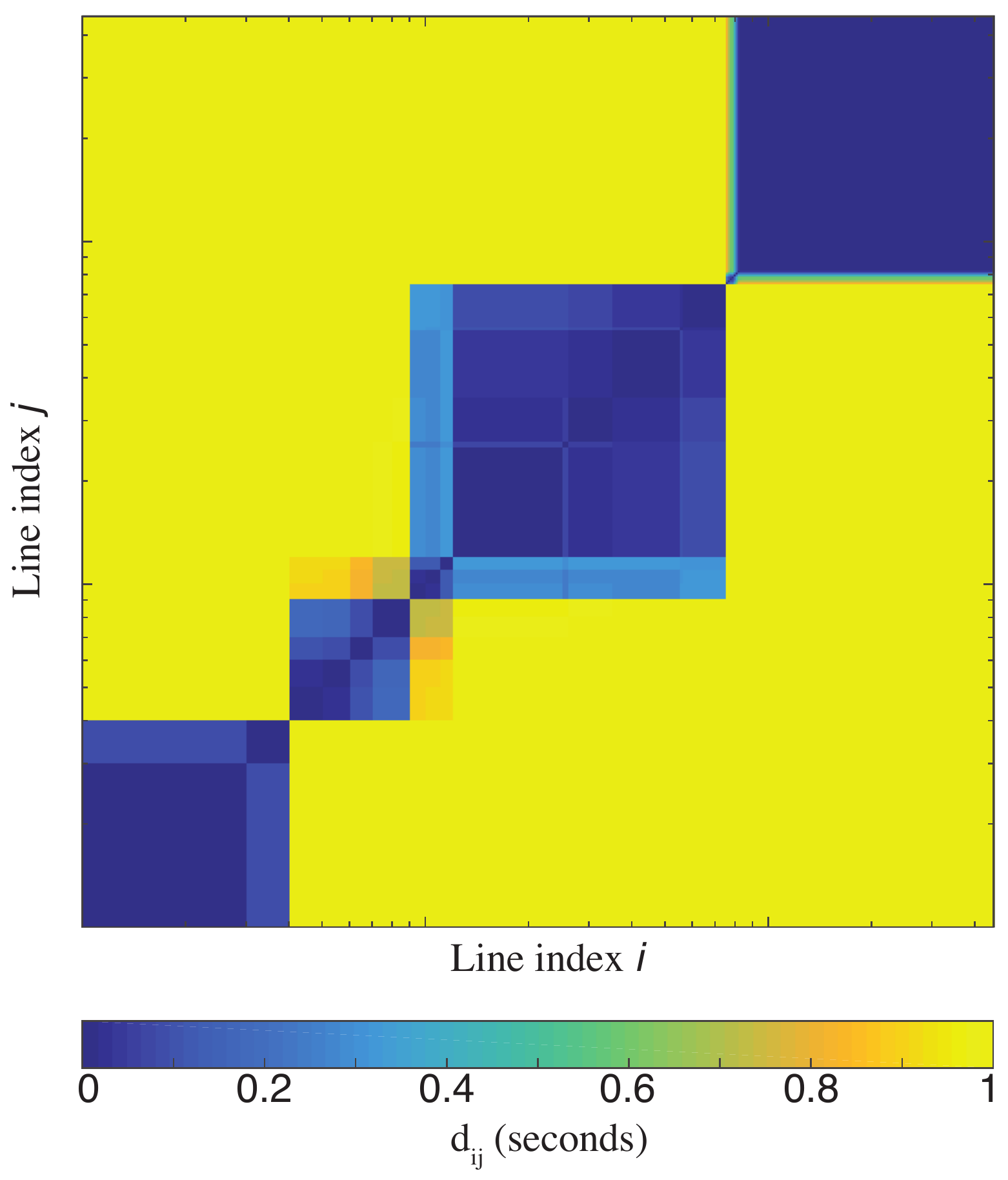}
\caption{We plot $d_{ij}$ as a function of the line indexes $i$ and $j$, where we have sorted the lines by their expected failure time. Blue indicates a small value of $d_{ij}$, while yellow indicates $d_{ij}\geq1$. The axes are on a log scale to show that successive failures result in (roughly) an  exponentially increasing number of line disconnections.  }\label{fig::dij}
\end{center}
\vspace{-0.5cm}
\end{figure}


For this parameterization, over $99\%$ of simulations resulted in failure in the chosen time horizon, where the condition for network failure is the isolation of all load nodes from the slack node. In our simulations, network failure occurred as a result of all lines having been disconnected from the slack. In less than $1\%$ of the simulations where the network did not completely fail, the network remained stable up to $t=T$, with around $60\%$ of the total load being served (consistently in the stable case $L(T)=0.58$).

Relabeling the lines by their ranked mean disconnection time, we plot $d_{ij}$ in Figure \ref{fig::dij}, computed from all trajectories that ended in network failure. The apparent structure shows large numbers of lines failing simultaneously (because of their disconnection from the network), partitioned into groupings. Between these groups, we can see that a waiting period before the subsequent group of lines fails. 
Non-overlapping squares indicate that one must wait for over 1 s  between successive disconnection events (as any $d_{ij}>1$ is colored yellow). We plot the data on a log scale to illustrate that the scale of these waves of disconnections grows exponentially as parts of the network become isolated from the slack node.

Running the hierarchical clustering algorithm described in Section \ref{sec::simulation} on this data gives five clusters that match the five blue boxes in the diagram, where the fourth cluster is only six lines and the fifth cluster corresponds to all remaining lines. Grouping the lines into these clusters, we can extract expected transition  times from the data to build up a picture of our progress toward total disconnection in a simulation. 

\begin{figure}[t]
\begin{center}
\includegraphics[trim={0cm 0 0cm 0},clip,width=0.45\textwidth]{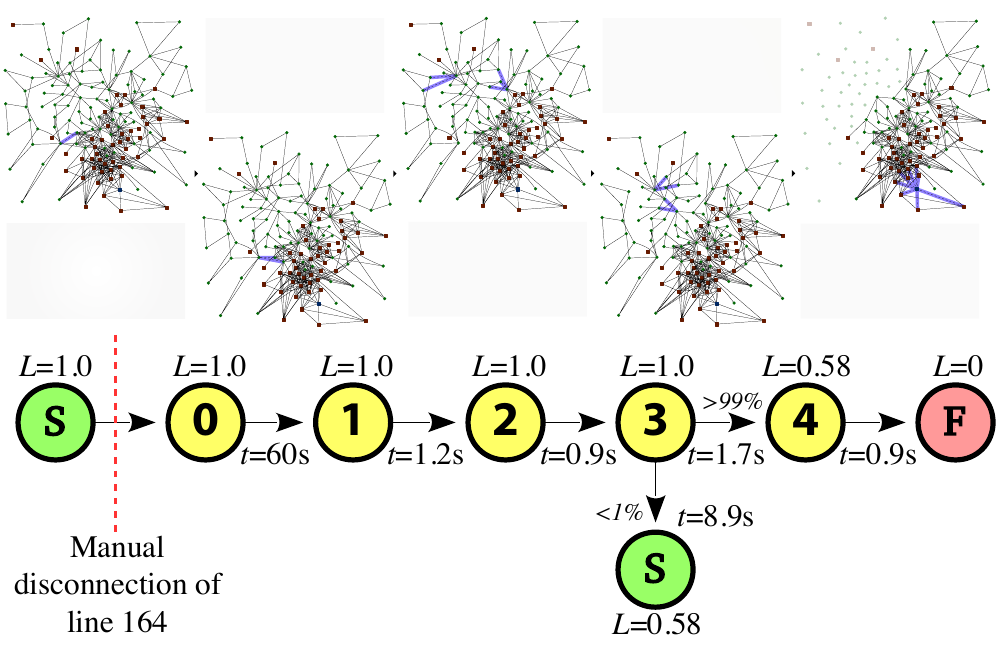}
\caption{The transition state flowchart for the model described in Section \ref{sec::results1}. Each node represents a state visible in Figure \ref{fig::dij}, with the expected time between transitions marked on each arrow and the expected load marked above each node. Diagrams for the network for states 0--4 are also pictured, with the failed lines marked in blue. }\label{fig::flowchart}
\end{center}
\end{figure}

If we consider the system entering a new state when a line disconnects from the next group, we can estimate the transition times between the states from the accumulated trajectory data. Beginning with the stable $\gamma=\mathbf{1}$ configuration, the subsequent state is entered after the initial disconnection of line 164 at $t=1$. The states displayed in Figure \ref{fig::flowchart} match the regions in Figure \ref{fig::dij}. In Figure \ref{fig::flowchart} we also plot the expected times for transitions between the states.

We conclude that our model indicates that the lines will fail in several clusters that can each be considered to fail almost simultaneously. This is apparent from the colors of the corner-pixels of each cluster's box in Figure \ref{fig::dij}, which gives $d_{ij}$ for the first and last line to fail in a particular cluster. This shows that we get a large chunk of disconnected lines within a small time, rather than a gradual disconnection. 

When in state 3 (as labeled in Figure \ref{fig::flowchart}) , almost all simulations go on to become unstable (ending in total disconnection), moving to state 4 in a mean time of $1.7$ s. However, some simulations (less than $1\%$) took far longer to leave state 3 and resulted in a configuration that remained stable over a long time. We can interpret this as the system exiting the well associated with this state at different points in space, where one exit (leading to state 4) is lower energy (and hence more probable) than the other. 

This example provides insight into an effective strategy setting relay thresholds in order to minimize the portion of the network ``lost'' through the action of protection. One could  lower the barrier for the exit point leading to this stable configuration, in order to increase the probability that we transition into the stable basin. Doing so amounts to making the relevant line (line 17)  disconnect automatically upon reaching state 3---purposefully sacrificing part of the network in order to save the $58\%$ load that is supplied.  Finding which line we need to disconnect in order to enact this strategy amounts to examining the failure pathway of the trajectories that correspond to this stable transition. In this case, we find that line 17 failing first is correlated with the stable transition in state 3.

\subsection{Failure times at higher thresholds} \label{sec::results2}
We now look at how changing the global threshold for relay tripping affects the expected time for total failure of the network. We vary the threshold between the $6.5\%$ value used in Section \ref{sec::results1} and $7.9\%$ of a line's susceptance value.   For each threshold value we run sixteen independent experiments to gather statistics. 
For experiments using the highest thresholds, line failures become significantly rarer after the initial disconnection; therefore,  accumulating good statistics on failure times requires a longer wall-clock time without enhanced sampling techniques.

We compare results from using the brute force strategy, where we simply sample independent trajectories in parallel, with results from using the ParRep method.
For the latter scheme, we run on ten Intel E5-2670 nodes, using $160$ parallel execution threads over MPI.   We use $T_\mathrm{dephase}=T_\textrm{decorr}=100$ s, for all the experiments, which gives the system enough time to decorrelate from an initial condition.  

We compare the results from experiments at fifteen different threshold values within our specified range. The qualitative path to failure for the network (see Figure \ref{fig::flowchart}) was not significantly different across the ranges of thresholds tested, but the time of the first failure after initial disconnection significantly increased as the threshold was raised. The reason is that subsequent failures cause a surge in line energy large enough that we see successive, rapid follow-on failures, whereas the initial failure  comes not from a surge but from the system wandering far enough away from $\xeq$ in a random walk.

\begin{figure}
\begin{center}
\includegraphics[trim={0cm 0 0cm 0},clip,width=0.4\textwidth]{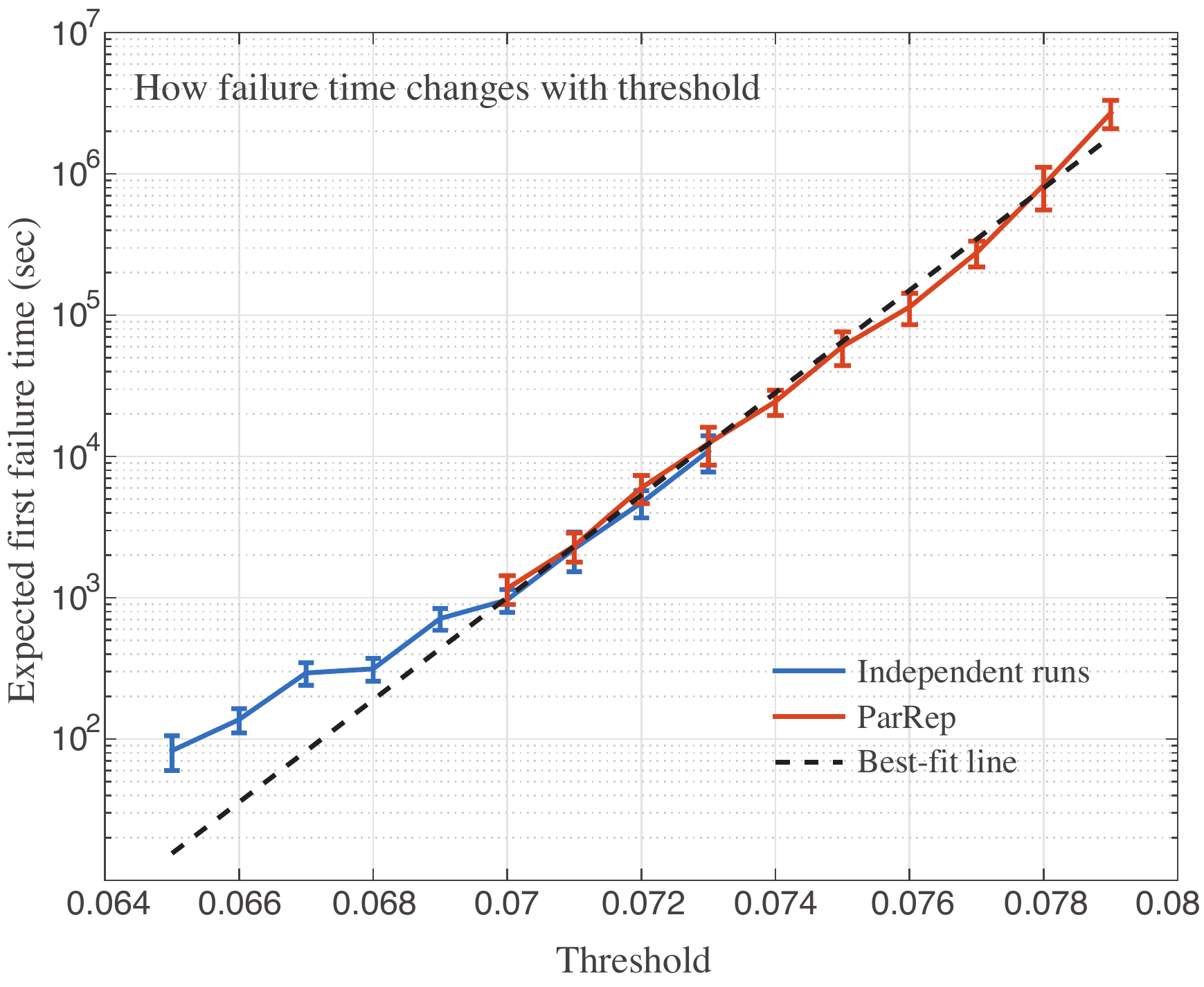}
\caption{The mean first failure of the power network as we change the global failure threshold, computed from the direct sampling (blue) and parallel replica (red) schemes. The two methods show good overlap, demonstrating an exponential relationship between failure time and failure threshold. }\label{fig::ftime}
\end{center}
\end{figure}

In Figure \ref{fig::ftime} we plot the mean time for first failure as we change the global threshold, with error bars computed over the sixteen runs. We run experiments using the direct sampling scheme up to the $7.3\%$ threshold and experiments using ParRep from $7.0\%$ onwards. The overlap in results from the repeated experiments demonstrates that the two methods have good agreement, falling within each other's error bars. The reason for not completing the entirety of the experiments with direct sampling is  the increase in wall-clock time: the total wall-clock time should scale linearly with the failure time of the system, and hence the wall-clock time at $7.9\%$ will be approximately $200\times$ larger than that at $7.3\%$. However, ParRep gives an approximately linear speedup as we use more parallel threads, allowing us to reach these longer timescales.


Our results suggest an exponential relationship between the failure time and threshold, although at lower thresholds the network takes longer than expected to fail. The reason may be that our initialization period is too short, so the system begins too close to $\xeq$ rather than being decorrelated. As the threshold is increased, this effect decreases because the initial decorrelation time becomes small relative to the exponentially increasing failure time.
Even a small increase in the line-tripping thresholds has a large impact on the stability of the network. From the results in Figure \ref{fig::ftime}, we can see that around a $0.3\%$ increase in the absolute  line threshold yields a factor of 10 increase in expected time to failure.  




\section{Conclusion} \label{sec::conclusion}
 In this article we have introduced and developed a novel framework for studying the evolution of dynamics in electrical power grids. By adding a stochastic term to the constant-energy dynamics the system explores state space independently of the initial condition provided and at a rate inferred from physical observations. We have demonstrated how statistics, including failure rates and paths toward cascade failure, can be gathered by solving the associated SDEs.

The approximate solution to these SDEs can be found by iterating a simple algorithm that, when tested, is found to be relatively cheap (compared with alternatives) and minimizes the error in statistics computed along the solution trajectory. Additionally we give an enhanced sampling algorithm that allows for significant improvements over direct sampling when a large amount of computing resources are available.

Numerical experiments show an exponential relationship between the line failure threshold and the failure of lines.

\section*{Acknowledgments}
This material was based upon work supported by the U.S. Department of Energy, Office of Science, Office of Advanced Scientific Computing Research (ASCR) under Contract DE-AC02-06CH11347. 
M. A. acknowledges partial NSF funding through awards FP061151-01-PR and CNS-1545046. The efforts of J. W. were also supported by ASCR through award DE-SC0014205.
We acknowledge the University of Chicago Research Computing Center for support of this work.

\bibliographystyle{plain}
\bibliography{main}

 \vspace{-0.15cm}
 \begin{flushright}
\scriptsize \framebox{\parbox{2.5in}{Government License:
The submitted manuscript has been created by UChicago Argonne, LLC,
Operator of Argonne National Laboratory (``Argonne").  Argonne, a
U.S. Department of Energy Office of Science laboratory, is operated
under Contract No. DE-AC02-06CH11357.  The U.S. Government retains for
itself, and others acting on its behalf, a paid-up nonexclusive,
irrevocable worldwide license in said article to reproduce, prepare
derivative works, distribute copies to the public, and perform
publicly and display publicly, by or on behalf of the Government. }}
\normalsize
\end{flushright}

\newpage

\appendices 
\section{Numerical discretization algorithms } \label{app::methods}
The numerical algorithms we use in Section \ref{sec::simulation} for discretizing \eqref{eqn::sdes} are given below, where our approximation for $x(t)$ is denoted $x_t$. $R_t$ is a vector of $3N$ i.i.d. unit normal random numbers, with $\E( R_t \cdot R_{s} )=3N\delta_{ts}$.
\subsection{Euler scheme}
See any relevant statistics textbook, e.g. \cite{milstein2004stochastic}, for a detailed discussion on the Euler and Heun discretizations.
\[
x_{t+h}= x_t - h (J-\epsilon S)\nabla H(x_t) + \sqrt{2{h\epsilon\tau S}}R_t
\]

\subsection{Heun's method}
\begin{align*}
y_t &= x_t - h (J-\epsilon S)\nabla H(x_t) + \sqrt{2{h\epsilon\tau S}}R_t,\\
x_{t+h} &= x_t - \frac h2 (J-\epsilon S)(\nabla H(x_t)+\nabla H(y_t)) + \sqrt{2{h\epsilon\tau S}}R_t
\end{align*}

\subsection{Structure preserving scheme}
We split $J=J_1+J_2$ where
\[
J_1 = \left[ \begin{array}{ccc} 0 & 0 & 0 \\ I_{l}+I_g & 0 & 0 \\ 0 & 0 & 0 \end{array} \right],\quad  J_2 = \left[ \begin{array}{ccc} 0 & -I_{l}-I_g & 0 \\ 0 & 0 & 0 \\ 0 & 0 & 0 \end{array} \right]
\]
Then  the algorithm is written in-situ as
\begin{align*}
\hat{x} &\leftarrow x_t \\
\hat{x} &\leftarrow \hat{x} + \frac h2 J_1\nabla H(\hat{x})\\
\hat{x} &\leftarrow \hat{x} + \frac h2 J_2\nabla H(\hat{x})\\
y &\leftarrow \hat{x} - h\epsilon S \nabla H(\hat{x}) + \sqrt{2h\epsilon S\tau}R_t,\\
\hat{x} &\leftarrow \hat{x} - \frac h2\epsilon S \left( \nabla H(\hat{x}) + \nabla H(y)\right) + \sqrt{2h\epsilon S\tau}R_t,\\
\hat{x} &\leftarrow \hat{x} + \frac h2 J_1\nabla H(\hat{x})\\
\hat{x} &\leftarrow \hat{x} + \frac h2 J_2\nabla H(\hat{x})\\
x_{t+h} &\leftarrow \hat{x}
\end{align*}

\subsection{LM scheme}
The details of this algorithm can be found in [4]. 
\[
x_{t+h}= x_t - h (J-\epsilon S)\nabla H(x_t) + \sqrt{\frac{h\epsilon\tau S}2}(R_t + R_{t+h})
\]

\end{document}